\documentclass[11pt]{article}
\usepackage{blindtext}
\usepackage[a4paper, total={6in, 8in}]{geometry} % 8.1in

\usepackage{graphicx}% Include figure files
\usepackage{dcolumn}% Align table columns on decimal point

\usepackage{ascmac}
\usepackage{bm}  % bol maththick vector
\usepackage{autobreak}
\usepackage[utf8]{inputenc}
\usepackage[T1]{fontenc}
\usepackage{mathptmx}

\begin{document}

\title{Molecular Dynamics Simulations of Ice and Methane Hydrate \\
by Means of the Rotation Coordinate TIP5P-Ewald Model}
% Tanaka and Papadopoulos

\author{Motohiko Tanaka \\
Innovative Energy Science and Engineering, Chubu University, Japan.}
%\altaffiliation

%\homepage{http://photon.isc.chubu.ac.jp/.}

\date{\today}% It is always \today, today,
\maketitle

%                                                                   without methane            
\begin{abstract}
Molecular dynamics simulations are utilized to study the microwave heating of ice and methane hydrate by the five-body rotation coordinate system with the TIP5P-Ewald model. The structure I of methane hydrate is constructed, and the ice or methane hydrate are exposed to microwave electric fields of 10 GHz.
Provisional methane hydrate of the normal density and a temperature of 273 K is dynamically unstable and collapses after some periods of irradiation. The period of a collapse time is $1.7 \times 10^{6} \tau $ and the temperature increases as $\Delta T \cong 61$ deg, with the external electric field $3 \times 10^{7} \rm{V/cm}$  (i.e. 0.3 V/\AA) and $\tau = 1 \times 10^{-14}$ s. 
For the methane hydrate of the temperature 193 K and the pressure 1 atm, the system is stable while it is heated under microwave irradiation. About the temperature of 273 K, high density methane hydrate becomes stable, whereas the density of 0.93 g/cm$^{3}$ is marginally stable while is heated when microwaves are present.   
In the microwave device of 1,000 V/cm and the pressure 1 atm, the ice points to 1 s in the 100\% microwave efficiency.

\vspace{0.2cm}
\noindent
{\small {\it 
Subjects: Chemical Physics (physics.chem-ph), \\ 
http://physique.isc.chubu.ac.jp/, http://www1.m4.mediacat.ne.jp/dphysique/, 
\\ 
This article utilizes molecular dynamics simulation of the rotation coordinate system and discusses various density and temperature emvironments of the ice and methane hydrate against microwave experiments. 
}}
\end{abstract}

%\maketitle

%\begin{quotation}
%The ``lead paragraph'' is encapsulated with the \LaTeX\ 
%\verb+quotation+ environment and is formatted as a single paragraph before the first section heading. 
%(The \verb+quotation+ environment reverts to its usual meaning after the first sectioning command.) 
%Note that numbered references are allowed in the lead paragraph.
%
%The lead paragraph will only be found in an article being prepared for the journal \textit{Chaos}.
%\end{quotation}

% Sections from 1,2,3,4
%1 Introduction
%2 Methodology of molecular dynamics by the TIP5P-Ewald model
% 2.1 General equations of motion and forces
% 2.2 Procedures of the translation and rotation coordinate system
%  2.2.1 The summary of the five-body water model
%  2.2.2 A la carte procedures of the five-body water mode
%3 Simulations of methane hydrate by the TIP5P-Ewald model
% 3.1 Simulation modeling}
% 3.2 Dependence of heating on the microwave fields
%  3.2.1 Dependence of provisional heating at the temperature 273 K
%  3.2.2 Dependence of heating at the temperature 273 K
% 3.3 Dependence of heating on high and medium densities
% 3.4 Dependence of heating on the temperature 193 K
%% 3.5 The microwave devices of the input power 1,000 V/cm}
%4 Summary
% Acknowledgment

%\end{document}

\section{\label{sec1}Introduction}
%\pagenumbering{arabic}

The natural gas resources of methane hydrate that are found in permafrost and the sea floor of the earth may be calling public attention \cite{Sloan}. 
Methane hydrate is a solid or liquid material and is a light electrolyte like ice. 
Either decompression or heating of methane hydrate layers can be applied to obtain methane from the sea floor. 
The production of methane from methane hydrate, however, makes about thirty times more climate warmings than carbon deoxide by greenhouse effects \cite{Pavlov, EPA}. 

Methane hydrate has a density of 0.91 $\rm{g/cm}^3$ at an atmospheric pressure and 0.95 $\rm{g/cm}^3$ for an elevated pressure. % around 50 atm. 
Methane hydrate dissociates to about 220 ml of methane gas against 1 g water at 1 atm and 273 K. It is stable at pressures higher than 0.1 MPa at 193 K and 2.3 MPa at 273 K. 
% I: (CH4)8(H2O)46, II: 136(H2O)16(CH4)8(C2H6)
There are three states of methane hydrates \cite{Sloan, matdata}. System I has 46 H$_2$O molecules that form $5^12$ and $5^{12} 6^{2}$ cages containing the guest molecules CH$_4$ and CO$_2$. System II has 136 H$_{2}$O molecules that form 5$^{12}$ and 5$^{12} 6^{4}$ tetragonal cages containing oxygen and other molecules. Both systems have cubic lattices. The hexagonal system H has 5$^{12}$, 5$^{12} 6^{16}$, and $4^{3} 5^{6} 6^{3}$ polyhedron cages of the C$_{6}$H$_{14}$ molecules, which exist as a hexagonal lattice. 

High-pressure experiments on methane hydrate have been performed using a diamond-anvil cell \cite{Hirai}. Experiments below the melting point of ice surveyed energetically for hydrates. The stability of hydrates in the thermodynamic instability of the ice Ih clathrate has been discussed \cite{stab-md}.
Many traditional equations of state have been utilized to describe thermo-physical properties and phase equilibrium \cite{Soave}.
The multi-scale phase diagram of the Gibbs-Helmholtz constrained equation of state for methane hydrate has been tabulated by density for given pressures and temperatures \cite{lucia2010,density-tab}. 

The diffusion coefficients and dielectric relaxation properties of water, i.e., the response of electric dipoles to a given initial impulse, have been studied theoretically \cite{yamag}.
The heating and diffusion of water under high-frequency microwaves and infrared electromagnetic waves have been investigated by molecular dynamics simulations using
elaborated point-charge models \cite{engl}.
Molecular dynamics simulation of the ice nucleation and growth process leading to water freezing has been executed \cite{matsu}.

Concerning the microwave heating of water, ice and saline solution, molecular dynamics simulations have been the first publication as microwave heating of water \cite{cite01}. 
They have shown that: (i) water in the liquid phase is heated by excitation of water electric dipoles, which is delayed from the microwave electric field, and absorbed the total microwave power; (ii) hot water gains significantly less heat than the water at room temperature because of smaller phase lags due to less friction; (iii) water in the ice phase is scarcely heated because the electric dipoles can not rotate due to the tightly hydrogen-bonded ice crystal; (iv) dilute saline solution gains significantly more heat than pure water because of the rapid heating of salt ions, especially that of the large salt ions Cl$^{-}$ and Na$^{+}$. 

Molecular dynamics employing the density functional method (DFT) to simulate the THz range of electromagnetic wave have been constructed \cite{cite02}.
They have shown, by the self-consistent atomic forces \cite{Siesta}, that: (i) liquid water molecules in the electric field has excited rotational motions, as water molecules in the cages can not make free translation motions; (ii) the electron energy is about twice the kinetic energy of the water molecules, which results from the forced excitation of the molecules by the electromagnetic THz external field.

%Regarding the methane hydrates as natural resources, there are several methods for collection from the sea floor though ${\rm CH}_4$ deposited in air might be very harmful to climate warmings. 

The goals of molecular dynamics simulations of the rotation coordinate system with the TIP5P-Ewald model are the followings:  
(i) how the CH$_{4}$-free ice behaves at 273 K against microwave irradiations,  
(ii) how methane hydrate is stabe at the low temperature at 193 K,  
(iii) how the density of the stable and unstable boundary of methane hydrate exists at the temperature at 273 K ?

The rest of this paper is organized as follows. The methodology of molecular dynamics with equations of motion and forces is written in Section \ref{sec2.1}, and the procedures of the five-body water model are given in Section \ref{sec2.2}. Modeling of present simulations is given in Section \ref{sec3.1}, and heating and collapse of methane hydrate at the temperture 273 K and the pressure 1 atm are shown in Section \ref{sec3.2}. 
Simulations of variable densities with constant temperture are shown in Section \ref{sec3.3}. 
% Error: temperature
The low temperature of 193 K with the pressure 1 atm is argued in Section \ref{sec3.4}.
The microwave devices of 1,000 V/cm are estimated in Section \ref{sec3.5}. 
A summary is provided in Section \ref{sec4}. 
The equations of long-range Coulombic interactions are explained in Appendix A.

%\end{document}

\section{\label{sec2}Methodology of molecular dynamics by the TIP5P-Ewald model}

\subsection{\label{sec2.1}{General equations of motion and forces}}

Crystal structures are fundamental aspects of solid-state physics \cite{Kittel}
which may be melted to liquid for elevated temperatures.
Molecules of ice and liquid water are within such categories. 
Four basic quantities in the CGS units system are used to derive the Newtonian equation of motion which are, (i) time $\tau = 1 \times 10^{-14} \ \rm{s}$, (ii) the length 1 \AA = $1 \times 10^{-8} \ \rm{cm}$, (iii) mass of water $M_{0} = 3.0107 \times 10^{-23} \ \rm{g}$, and  (iv) electronic charge $e = 4.8033 \times 10^{-10} \ \rm{esu}$ \ ($1.6022 \times 10^{-19} \ \rm{C}$ in the international units system). 
Then, one has the three-dimensional electrostatic version of equations of motion which are written as \cite{cite01,cite03}, 
\begin{eqnarray}
M_{j} \frac{d\bm{V}_{j}}{dt} &=&  -\nabla  \left\{ 
\Phi_{F} (\bm{R}_{j}) + \sum_{k=1}^{N \star} \left[ A/r_{jk}^{12} - B/r_{jk}^{6} \right] \right\}  + q_{j}E \sin \omega t \ \hat{x}, 
\label{eq1} \\
\frac{d\bm{R}_{j}}{dt} &=& \bm{V}_{j}.
\label{eq2}
\end{eqnarray}
%
%that define the equations of motion.
The first term of the right-hand side of Eq.(\ref{eq1}) is the Coulombic potential $\Phi_ {F}(\bm{R}_{j})$, and the second term is the Lennard-Jones potential. Here, $\bm{R}_{j}$ and $\bm{V}_{j}$ are the position and velocity of $j$-th molecule, respectively, $M_{j}$ and $q_{j}$ are the mass and charge, respectively, $t$ is the time, and $\nabla$ is the space derivative. The quantity $\bm{r}_{jk}= \bm{r}_{j} -\bm{r}_{k}$ is the particle spacing between the $j$-th and $k$-th atoms. 
Also, the symbol $\sum^{\star}$ of Eq.(\ref{eq1}) means to take the summation of the oxygens, %\bigtriangleup
and the symbol $\sum^{\bullet}$ of Eq.(\ref{Shortrg}) in the paragraph later is that of the hydrogens.  
$A$ and $B$ are the coeffcients of Lennard-Jones potential, respectively, with $A= 4 \varepsilon \sigma^{12}$ and $B= 6 \varepsilon \sigma^{6}$ where $\varepsilon$ is the depth of the potential well and $\sigma$ is the distance at which the particle potential energy bocomes zero. 
%where r is the distance between two interacting particles, ε is the depth of the potential well (usually referred to as 'dispersion energy'), and σ is the distance at which the particle-particle potential energy V is zero

The external electric field $E$ points to the $x-$direction and has the sinusoidal form $\sin \omega t$ where the frequency $f$ is $\omega= 2 \pi f$.
The time step is $\Delta t= 0.025 \tau $ (i.e., $2.5 \times 10^{-16} \rm{s}$).
In a time marching fashion, the current step of $\bm{R}_{j}$ and $ \bm{V}_{j}$ is forwarded to the next time step. When a sufficient amount of time has elapsed, one analyzes the time development. 

To represent the crystal system with high accuracy, one has to separate the Coulombic forces $\bm{F}(\bm{R}_{j})= - \nabla \Phi_{F}(\bm{R}_{j})$ that occur in the short-range and  long-range interactions \cite{Ewald, Kolafa, Frenkel},
\begin{eqnarray}
\bm{F}(\bm{R}_{j}) &=& \bm{F}_{SR}(\bm{R}_{j}) +\bm{F}_{LR}(\bm{R}_{j}).
\end{eqnarray}
The short-range interactions are written as,
\begin{eqnarray}
\bm{F}_{SR} (\bm{R}_{j}) &=& \sum_{k=1}^{N \bullet}  q_{j}q_{k} \Bigl[ \Bigl( \frac{\rm{erfc}(\it{r}_{jk})}{r_{jk}} +\frac{2\alpha}{\sqrt \pi} \Bigr) \exp(-(\alpha r_{jk})^2)/r_{jk}^{2} \Bigr] \bm{r}_{jk},
\label{Shortrg} 
\end{eqnarray}
where the Gauss complimentary error function is
\begin{eqnarray}
\rm{erfc}(\it{r})= \frac{2}{\sqrt \pi} \int_{r}^{\infty} \exp(-t^{2}) dt. 
\label{contgaus}
\end{eqnarray}
The $\alpha$ value, a minimization factor, is discussed later.

A primary factor in the long-range interactions is the charge density, $\rho(\bm{R}_{j})= \sum_{k} q_{k}S(\bm{R}_{j}-\bm{R}_{k})$, which is the near-site grid sum with $S(\bm{0})= 1$, $S(\infty)\rightarrow 0$. Then, the grid summation is converted to the k-space by a Fourier transform $FT^{-1}[...]$.
Here, $\rho(\bm{R}) \rightarrow \rho_{k}(\bm{k})$ with $\bm{k}= 2 \pi \bm{n}/ \it{L}$, $n$ the integers $\ge 0$, and $L$ the length. The inverse Fourier transform to return to the coordinate space is executed by the folding operations $FT[...]$,
\begin{eqnarray}
 \bm{F}_{LR}(\bm{R}_{j}) &=& -FT \Bigl[ i \ q_{j} 
   (dn(n_x),dn(n_y),dn(n_z)) G(n_x,n_y,n_z) \ \rho_{k}(\bm{k}) \Bigr] , 
   \label{Longrg1}\\
  && dn(n_{\gamma})= n_{\gamma} - dnint(n_{\gamma}/M_{\gamma}) M_{\gamma}
 \hspace*{0.3cm} (\gamma= x,y,z).
\label{Longrg} 
\end{eqnarray}
The expressions for the $G(n_x,n_y,n_z)$, $\bm{K}(n_x,n_y,n_z)$ and $\Delta(n_x,n_y,n_z)$ functions are given in Appendix A. The $\alpha$ value is determined by minimizing the errors of both the short-range and long-range interactions of the electric fields \cite{Deserno}. The value is $\alpha=0.203$ for the total number of $3^3$ methane hydrates. 

The five-body molecules are used for water which is known as the TIP5P-Ewald model in the periodic boundary condition.
The asymmetric pyramid of 5-body water molecules is specified to assign positive two hydrogens $ q_{H}= 0.241 e$ and negative two hydrogens $ q_{L}= -0.241 e$ with $e$ the electron charge. 
The angle and bond, respectively, are $\psi_{1} = 104.52^\circ$, $r= 0.9572$ \AA \ for the $H$ sites and $\psi_{2} = 109.47^\circ$, $r= 0.7$ \AA \ for the dummy $L$ sites whose dynamics motion is not counted.
The fifth oxygen atom $ q_{O}= 0 $ is to correlate with adjacent molecules using the Lennard-Jones potential $ \Psi (r)= A/r^{12} -B/r^{6} $. The factors of the TIP5P-Ewald water model are $ A= 3.8538 \times 10^{-8} $ erg \AA$^{12}$ and $ B= 4.3676 \times 10^{-11} $ erg \AA$^{6}$ \cite{Rick}. 
%For methane case, $\epsilon_{ij}$ and $\sigma_{ij}$ are used as the two-particle interaction potential and the size of the CH$_{4}$ molecule, respectively. 
%The classical mechanics and related topics are written in the Goldstein's book \cite{Goldstein}. 

\subsection{\label{sec2.2}Procedures of the translation and rotation coordinate system}

\vspace{-0.2cm}
\subsubsection{\label{sec2.2a}The summary of the five-body water model}

\noindent
A) Five sites are one oxygen of $O$ site, hydrogens of $H_{1}$ and $H_{2}$, and 
negative hydrogen virtual sites of $L_{1}$ and $L_{2}$. 
Their charges are $0$, $0.241 e$,  $0.241 e$, $-0.241 e$ and $-0.241 e$, respectively. The $L_{1}$ and $L_{2}$ are called the dummy sites.

\noindent
B) Separate the translational coordinate $\bm{R}_{j}$, $\bm{V}_{j}$ for molecules ($j= 1,N/5$) from the rotation coordinate $\bm{r}_{i}=  (x_{i}, y_{i}, z_{i})$ of atoms ($i= 1,N$) for the five sites.
The separation is done at the starting step only;
once determined at $t=0$, they become constant in time.

\noindent
C) For the rotation coordinate, the half time step for the molecules is first executed for a predictor step, and the full time step is made in advancement of time for a corrector step.

\noindent
D) Before the end of the cycle, the forces are calculated by atomic positions. The dummy sites at $L_{1}$ and $L_{2}$ are obtained  trigonometrically by vector products of $O, H_{1}$ and $H_{2}$ sites.

\noindent
E) After correction of quaternions, the kinetic and Coulombic energies are calculated, and go to the beginning of the cycle. 
The leap-frog method is used for the plasmas and water.

\vspace{-0.2cm}
%\vspace{0.2cm}
\subsubsection{\label{sec2.2b}A la carte procedures of the five-body water model}

Each step of the simulation cycle of five-body molecules is written, which is 
(i) Read positions $\bm{r}_{i}$ and quaternions $(e_{0},e_{1},e_{2},e_{3})$ for the first step,
(ii) translational motion (Step 1), 
(iii) rotational motion (Steps 2-4), and
(iv) the addition of the fields (Steps 5-8). 

%\vspace{0.1cm}
\noindent
(0) An extra procedure is made only at the first time step.
Read positions $\bm{r}_{i}$ from the file by $ ``read(17) \ x_{i},y_{i},z_{i} "$ with 
$i= 1,2,3,6,7,8,..$. Also, read the quaternions $\bm{q}_{j}$ from the file by 
$ ``read(30) \\ e_{0j},e_{1j}, e_{2j}, e_{3j} ``$ ($j=1,N/5$).
%                must \\

\noindent
(1) The position $\bm{R}_{j}$ and the velocity $\bm{V}_{j} $ of each molecule ($j=1,N/5$) are advanced by summation over five sites of forces $\bm{F}_{k}$ $(k=1,N)$ for the translational motion,
\begin{equation}
d\bm{V}_{j}/dt= (1/m_{j})\sum_{k=1}^{5} \bm{F}_{k}, \ \ 
d\bm{R}_{j}/dt= \bm{V}_{j}.
\end{equation}

\noindent
(2) The rotation coordinate of steps 2) to 5) are made for half a time steps $\Delta t_{1}=\Delta t/2$ by prediction, and the next time for a full time step $\Delta t_{2}= \Delta t$ by correction.
The angular momentum of rotational motion is calculated at a time step $\Delta t_{1}$ or 
$\Delta t_{2}$ by summation over the torque of five sites,
\begin{equation}
 d\bm{L}_{j}/d t_{n}= \sum_{k=1}^{5}(y_{k}F_{k}^{z} -z_{k}F_{k}^{y},
z_{k}F_{k}^{x} -x_{k}F_{k}^{z}, x_{k}F_{k}^{y} -y_{k}F_{k}^{x}).
\end{equation}

\noindent
(3) The angular frequency $\omega_{j,\alpha}$ is connected to the angular mementum 
and inertia moment $Im_{j,\alpha}$ with $\alpha= x,y,z$ and the matrix 
$A_{\alpha,\beta}$ is,
\begin{equation} 
\omega_{j,\alpha}= (A_{\alpha 1}L_{x} +A_{\alpha 2}L_{y} +A_{\alpha 3}L_{z})/
Im_{j,\alpha},
\label{omegbyA}
\end{equation}
\begin{equation}
\begin{matrix}
A_{11}= e_{0}^2 +e_{1}^2 -e_{2}^2 -e_{3}^2, &
A_{12}= 2(e_{1}e_{2} +e_{0}e_{3}),&
A_{13}= 2(e_{1}e_{3} -e_{0}e_{2}),\\
\hspace{-0.34cm}
A_{21}= 2(e_{1}e_{2} -e_{0}e_{3}),&
A_{22}= e_{0}^2 -e_{1}^2 +e_{2}^2 -e_{3}^2, &
A_{23}= 2(e_{2}e_{3} +e_{0}e_{1}),\\
\hspace{-0.3cm}
A_{31}= 2(e_{1}e_{3} +e_{0}e_{2}),&
A_{32}= 2(e_{2}e_{3} -e_{0}e_{1}),&
\ \ \ A_{33}= e_{0}^2 -e_{1}^2 -e_{2}^2 +e_{3}^2. 
\label{matrixA}
\end{matrix}
\end{equation}

\noindent
(4) The time derivative of quaternions $\bm{q}_{j}= (e_{0},e_{1},e_{2},e_{3})$ is given 
by the angular frequency by,
\begin{equation} 
d\bm{q}_{j}/dt_{n}= (1/2) \Delta t_{n} 
\begin{pmatrix}
-e_{1} \omega_x -e_{2} \omega_y -e_{3} \omega_z \\
e_{0} \omega_x -e_{3} \omega_y +e_{2} \omega_z \\
e_{3} \omega_x +e_{0} \omega_y -e_{1} \omega_z \\
-e_{2} \omega_x +e_{1} \omega_y +e_{0} \omega_z
\end{pmatrix}
.
\label{quats1}
\end{equation}

\noindent
(5) Get the rotation matrix $ A_{\alpha \beta}(e_{0},e_{1},e_{2},e_{3})$ in half a time steps 
for the prediction. Go back and repeat Step 2 - Step 4 for a full time step for the correction step, and go to Step 6.

\noindent
(6) The three sites $\bm{r}_{i}$ and the position $\bm{R}_{j}$ are connected by,
\begin{equation}
  \bm{r}_{i}= \bm{R}_{j} +
     \begin{pmatrix}
     A_{11}  & A_{21}  & A_{31} \\
     A_{12}  & A_{22}  & A_{32} \\
     A_{13}  & A_{23}  & A_{33} % \nonumber
     \end{pmatrix}
     \begin{pmatrix}    
     x_{i} \\ y_{i} \\ z_{i} % \nonumber
     \end{pmatrix}
     .
\end{equation}
The positions of dummy sites $L_{1}$ and $L_{2}$ are calculated by the vector product from the known three sites of $O$, $H_{1}$ and $H_{2}$.

\noindent
(7) The forces for atomic positions are calculated from the Coulombic and Lennard-Jones potentials using the five sites.

\noindent
(8) Correction to the unity normalization of quaternions is made at every 10 steps. 
Go to the new time step of Step 1.

On the conversion from the Cartesian to the rotation at the initial time step, the standard procedures are the followings \cite{cite04}. \\
%        [ɑɪsά səlìːz] liːzが強く
(1) The isosceles triangle $\triangle O H_{1} H_{2}$ with $O$ on the top and $H$'s on the bottom is rotated by three times to be the angles $(\phi, \theta, \psi)$, with $0 \le \theta \le \pi$, $0 \le \phi \le 2 \pi$ and $0 \le \psi \le 2 \pi$. \\
(2) The angles $(\theta,\phi,\psi)$ are transformed to be the initial step quaternion $(-e_{0},-e_{1},e_{2},-e_{3})$, where a set of quaternion $(e_{0},e_{1},e_{2},e_{3})$ is defined by 
\begin{equation}
\begin{matrix}
e_{0}= cos(\theta/2) cos((\phi+\psi)/2), \\
e_{1}= sin(\theta/2) cos((\phi-\psi)/2), \\
e_{2}= sin(\theta/2) sin((\phi-\psi)/2), \\
e_{3}= cos(\theta/2) sin((\phi+\psi)/2).
\end{matrix}
\label{quats}
\end{equation}
The matrix $A_{ij}$ by quaternions was shown in Eq.(\ref{matrixA}).
It is noted that the equations Eq.(\ref{matrixA}) and Eq.(\ref{quats1}) differ by the choice of quaternions Eq.(\ref{quats}), then this article is followed by Goldstein's definition \cite{Goldstein}.
The use of quaternions does not have singularity unlike the Euler representation.

A time step is important and it will be $\Delta t= 0.025 \tau$ or less by the time advancing scheme; otherwise the explicit code will be inaccurate or go to overflow.

\begin{table} %[h]
\caption{The run series of Runs A's and AR's, fixed density ($\rm{g/cm}^3$), electric field (V/\AA), microwave heating rate in the linear phase ($W_{0}/\tau$) with $W_{0}$ the initial kinetic energy of methane hydrate, the run time before the collapse, and guest molecules.
%The kinetic energy increases and collapses for all Run A's runs.
%Run A2R$^{*}$ is the ice and methane in the nonlinear stage with $\Delta T \cong 73$ deg at $t=1.3 \times 10^{5} \tau$, and is collapsed at $t=1.9 \times 10^{5} \tau$.
%%The heating rate of Run A3R$^{*}$ is $1.7 \times 10^{-5} \rm{W_0}/\tau$ and $T \cong 333$ in the linear phase, and collapses at $t=2.9 \times 10^{4} \tau$. 
%
(The time axis in the abscissa starts from null in Fig.\ref{tip580_Energy} later, while the $t>0$ part is executed with the microwave irradiation as used in Table \ref{Table-1}$^{*}$. The $t<0$ part is for a dry run without microwaves.)
}
% Subtract: a time rate $\Delta W_{dr}/\Delta \tau= 1.6 \times 10^{-7} W_{0}/\tau$ 
%
\label{Table-1}
  \centering 
  \vspace{-0.2cm}
  \begin{tabular}{clllll} 
\hspace{0.3cm} \\ \hline
 \hspace{0.2cm} series & density & $E$ & heating rate$^{*}$ & run from $t=0$ & guest molecules \hspace{0.1cm} \\ \hline
%
% noositaaru
 A1$^{*}$ & $0.912 \rm{g/cm}^3$ & 0.30 V/\AA & $3.0 \times 10^{-7} W_{0}/\tau$ & $1.7 \times 10^{6} \tau$ \  collapsed &  $\rm{CH}_4$ % 1.7
\\ 
% A2$^{*}$ & $0.912 \rm{g/cm}^3$ & 0.20 V/\AA & $1.1 \times 10^{-7} W_{0}/\tau$ &  - & $\rm{CH}_4$
% A2: t=19,000
 A2$^{*}$ & $0.912 \rm{g/cm}^3$ & 0.35 V/\AA & $1.7 \times 10^{-6} W_{0}/\tau $ & $2.3 \times 10^{5}\tau$ \ collapsed & $\rm{CH}_4$ 
% A2 1.3 and 2.3
%
% naoshita
%%A3$^{*}$ & $0.912 \rm{g/cm}^3$ & 0.40 V/\AA & $1.6 \times 10^{-5} W_{0}/\tau$ & $1.3 \times 10^{4} \tau$ \ collapsed & $\rm{CH}_4$ % 6.3
% A3 1.6 and 1.3
\\ \hline
%% New but need a time !!  period=7.0 negative !
 A0R$^{*}$ & $0.912 \rm{g/cm}^3$ & 0.10 V/\AA & $4.1 \times 10^{-8} W_{0}/\tau$ & $1.5 \times 10^{5}\tau$ \ running & $\rm{CH}_4$ \\ % free CH4 5.1d-07
 A1R$^{*}$ & $0.912 \rm{g/cm}^3$ & 0.30 V/\AA & $5.0 \times 10^{-7} W_{0}/\tau$ & $1.4 \times 10^{6}\tau$ \ collapse? & $\rm{CH}_4$  
 % flat: 14-37 periods 1.9d-7; T0=294 K (correct), dT= 21 deg, linear
 % 2.5d-7; T0= 308 K  T1= 35 deg  started to nonlinear 
 % 2.8d-7, T0= 318 K  3.3d-7, T0= 332 K  1.12d6 3.6d-7 T0= 346 K  4.0d-7 1.25d6 358 K
 % 1.24d6 4.0d-7 T0= 362 K    1.33d6 368 K 4.2d-7  collapse start !  1.37d6 383 K 4.7d-7
 % 4.8d-7 T0= 387 K 1.40d6   5.0d-7 392 K 1.43d6
 \\
%                                                                       8.5 \times 10^{-8} tt=15 period 
 %
 A2R$^{*}$ & $0.912 \rm{g/cm}^3$ & 0.35 V/\AA & $1.9 \times 10^{-6} W_{0}/\tau$ & $1.8 \times 10^{5}\tau$ \ collapsed & $\rm{CH}_4$ 
%\\
%
%%A3R$^{*}$ & $0.912 \rm{g/cm}^3$ & 0.40 V/\AA & $1.8 \times 10^{-5} W_{0}/\tau$ & $3.0 \times 10^{4}\tau$ \ collapsed & $\rm{CH}_4$
% A3R$^{*}$ & $0.912 \rm{g/cm}^3$ & 0.45 V/\AA & $1.5 \times 10^{-4} W_{0}/\tau$ & $1.0 \times 10^{4}\tau$ \ collapsed & $\rm{CH}_4$
 \hspace{0.3cm} \\ \hline  
  \end{tabular}
\end{table}

\section{\label{sec3}Simulations of methane hydrate by the TIP5P-Ewald model}
\subsection{\label{sec3.1}Simulation modeling}

To represent electrostatic effects for water molecules, one may adopt the three-body model  \cite{Andersen,Berendsen}, or various four-body to five-body models \cite{Jorgensen}.
Although the three-body model has a lower temperature problem than the real one, 
it has a stable symplectic integrator scheme in the coordinate space \cite{Kang,forest,okabe}.
The microwave heating and collapse of methane hydrate were studied using the three-body SPC/E model \cite{TSN}. 
The five-body models are currently widely used, but there occurs a unrealistic drift to increase 
the kinetic energy with a time rate $\Delta W_{dr}/\Delta \tau= 1.6 \times 10^{-7} W_{0}/\tau$ for the null external electric field. 
%                              ++++++
The kinetic energies of Run A1 to Run B's in Table \ref{Table-1} and Table \ref{Table-2} are subtracted from the tilted baselines. 
% +++++++++

The initial structure of methane hydrate is installed by the Genice program on the Linux system \cite{cite04,cite31}. 
The size of structure I methane hydrate has about 12  \AA \ as the crystal structure.  
%
%\footnote{                                                    ++++++
In the numerical simulations, the density of 0.7898 $\rm{g/cm}^{3}$ is for the normal density of $\rm{H}_{2} \rm{O}$.
The extra molecules $\rm{C} \rm{H}_{4}$ are added to make the total density of 0.912 $\rm{g/cm}^{3}$ or other molecules like $\rm{C} \rm{O}_{2}$.
% ++++  
A total of 3$^3$ methane hydrate in the three-dimensional water system makes the overall methane hydrate molecules.  
%                                                               ******
It is 36.01 \AA \ in the normal density hydrate Run A1, 
that of high density Run B1 is 35.51 \AA \ and Run B2 is 35.79 \AA , respectively, and  that of medium density Run B3 is 35.94 \AA . 
%, and that of low density Run B4 is 36.63 \AA.

The Lennard-Jones potential parameter for water was argued in Sec.\ref{sec2.2}. 
A system of guest molecules of $\rm{CH}_4$ which are the united atoms 
is used for the Lennard-Jones potential in the second term on the 
righthand side of Eq.(\ref{eq1}). 
$\epsilon_{ij}$ and $\sigma_{ij}$ are the two-particle interaction potential and the radius of the molecule, respectively, and the potential coefficients are $\epsilon_{CH_4}=0.39 \ \rm{kJ/mol}$ and $\sigma_{CH_4}=3.82$ \AA\ \cite{AV}.
%, and those for carbon dioxide are $\sigma_{CO_2}=4.00$ \AA \ and $\epsilon_{CO_2}=0.53 \ \rm{kJ/mol}$, respectively.
%
In order to perform the molecular dynamics simulation, the frequency is 
set to 10 GHz ($f= 2 \pi \times 10^{4} /\tau$).

The external electric field is changed for $E= 0.30 \rm{V}$/\AA\ of Run A1 
and $E= 0.35 \rm{V}$/\AA\ for Run A2 
in normal density of $0.912 \rm{g/cm}^3$ and the temperature of 273 K.       
Those for $E=$ 0.10 V/\AA \ of Run A0R and $E=$ 0.30 V/\AA \ for Run A1R are for the ice and methane simulations.
The external electric field $E= 0.35 \rm{V}$/\AA\ is done for high density of $0.963 \rm{g/cm}^3$ in Run B1, $0.941 \rm{g/cm}^3$ in Run B2, and $0.930 \rm{g/cm}^3$ in Run B3. 
The methane hydrate at the temperature of 193 K and the pressure 1 atm is simulated in Run C1.

%Fo.r numerical execution of the molecular dynamics simulations, a Fujitsu FX100 Supercomputer (52 ranks$\times$16 processors, 8 thread) is utilized. The computations took $1.57 \times 10^{-2}$ s per time step, and thus the run of $5.5 \times 10^6$ time steps requires 24 hours of computation. 
 
%\newpage
\vspace{0.3cm}

\begin{figure}[t]
\centering
 \vspace{0.3cm}
  \includegraphics[width=5.5cm]{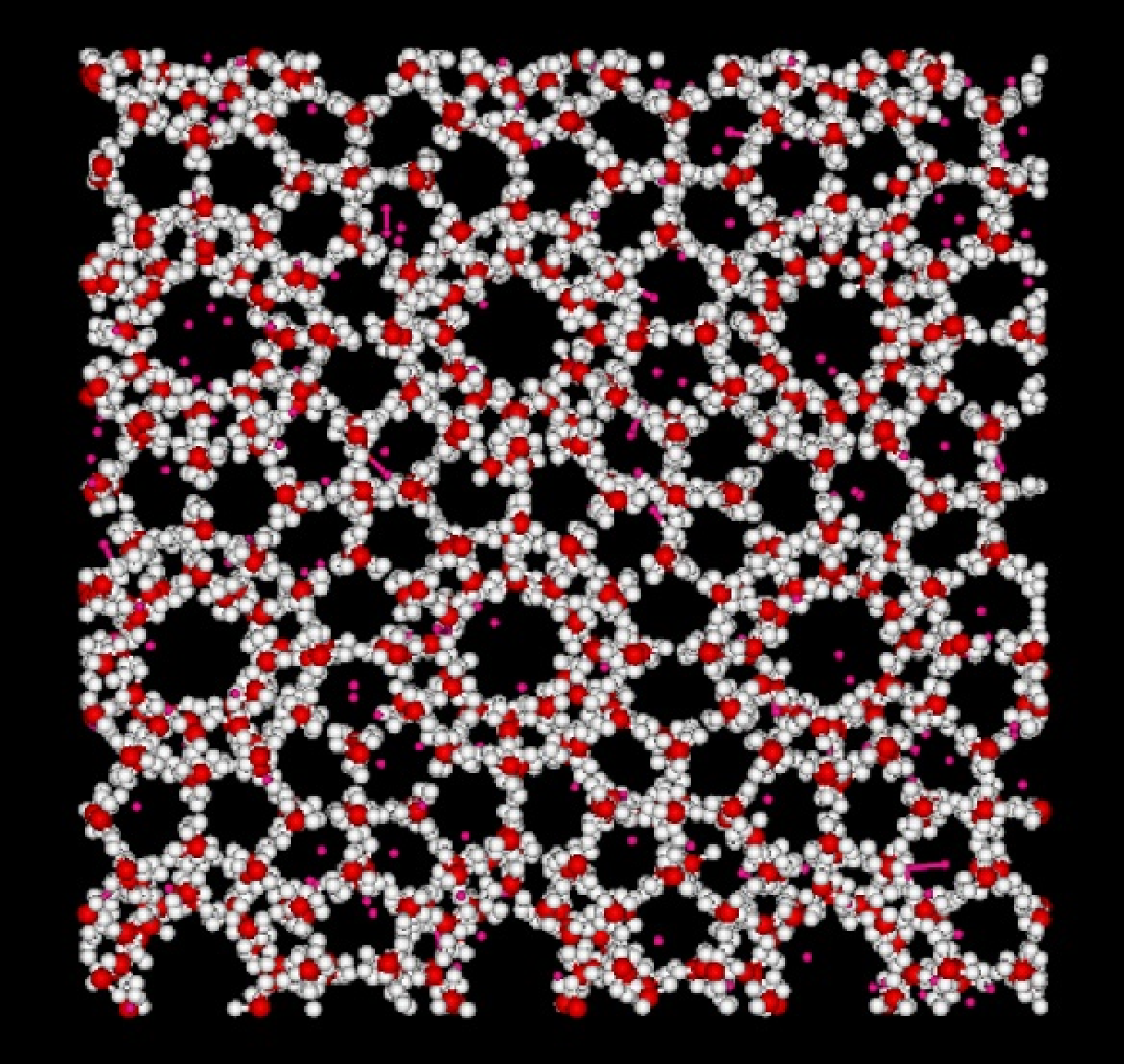}
\hspace*{0.2cm}
  \includegraphics[width=5.45cm]{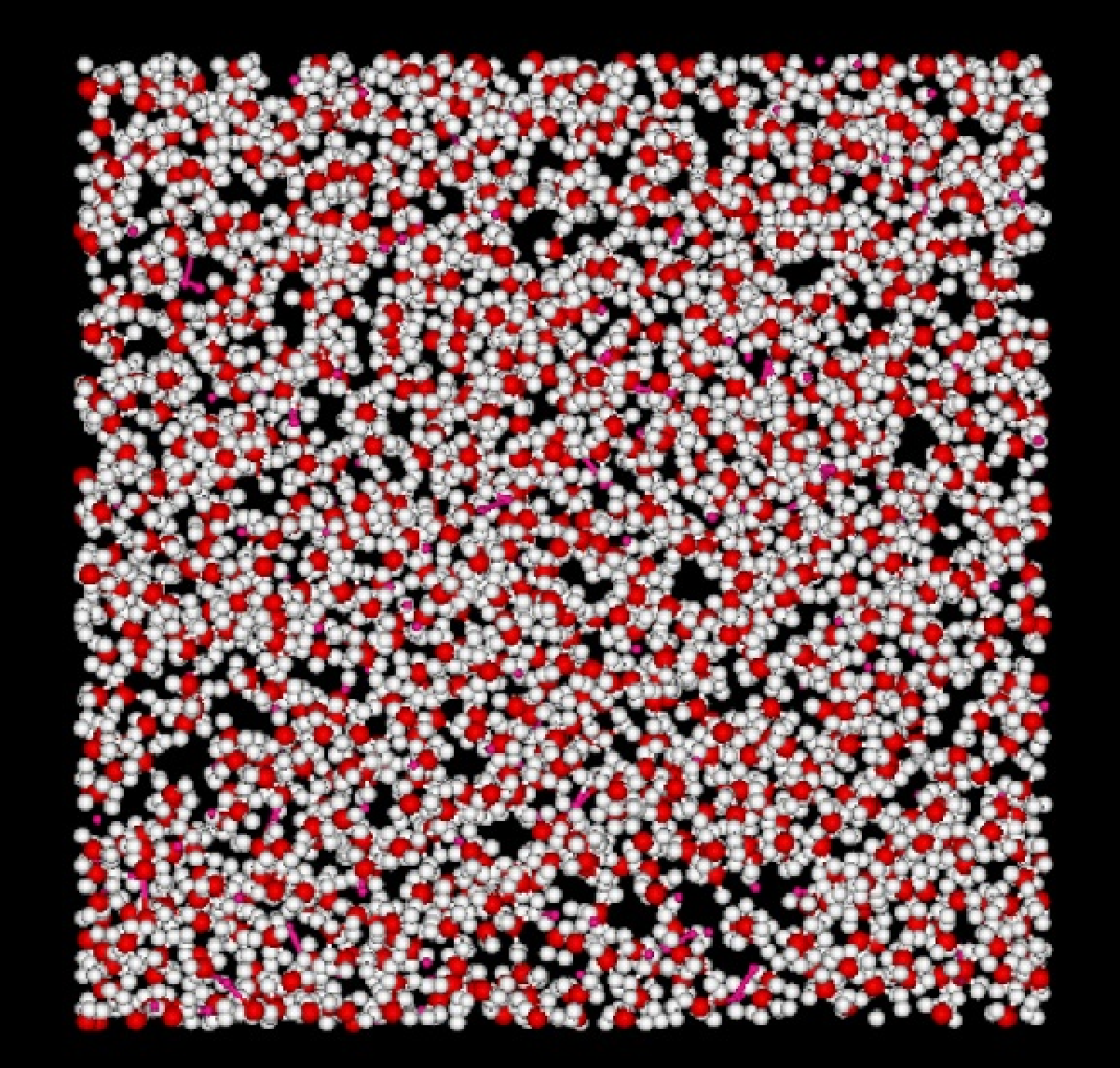}

\caption{
(Left) Methane hydrate of the H (white), O (red), and CH$_4$ (small red) molecules having density of 0.912 g/cm$^3$ and the external electric field $E= 0.30$ V/\AA\ before collapse at the time $ t = 1.60 \times 10^{6} \tau $,  
(Right) Distortion of the methane hydrate in liquid under an applied microwave field observed at $ t = 1.65 \times 10^{6} \tau $. % 1.70
}
\label{tip580_scat} % {mhr023plot}
\end{figure}

\begin{figure}
\centering
  \includegraphics[width=11.0cm,clip]{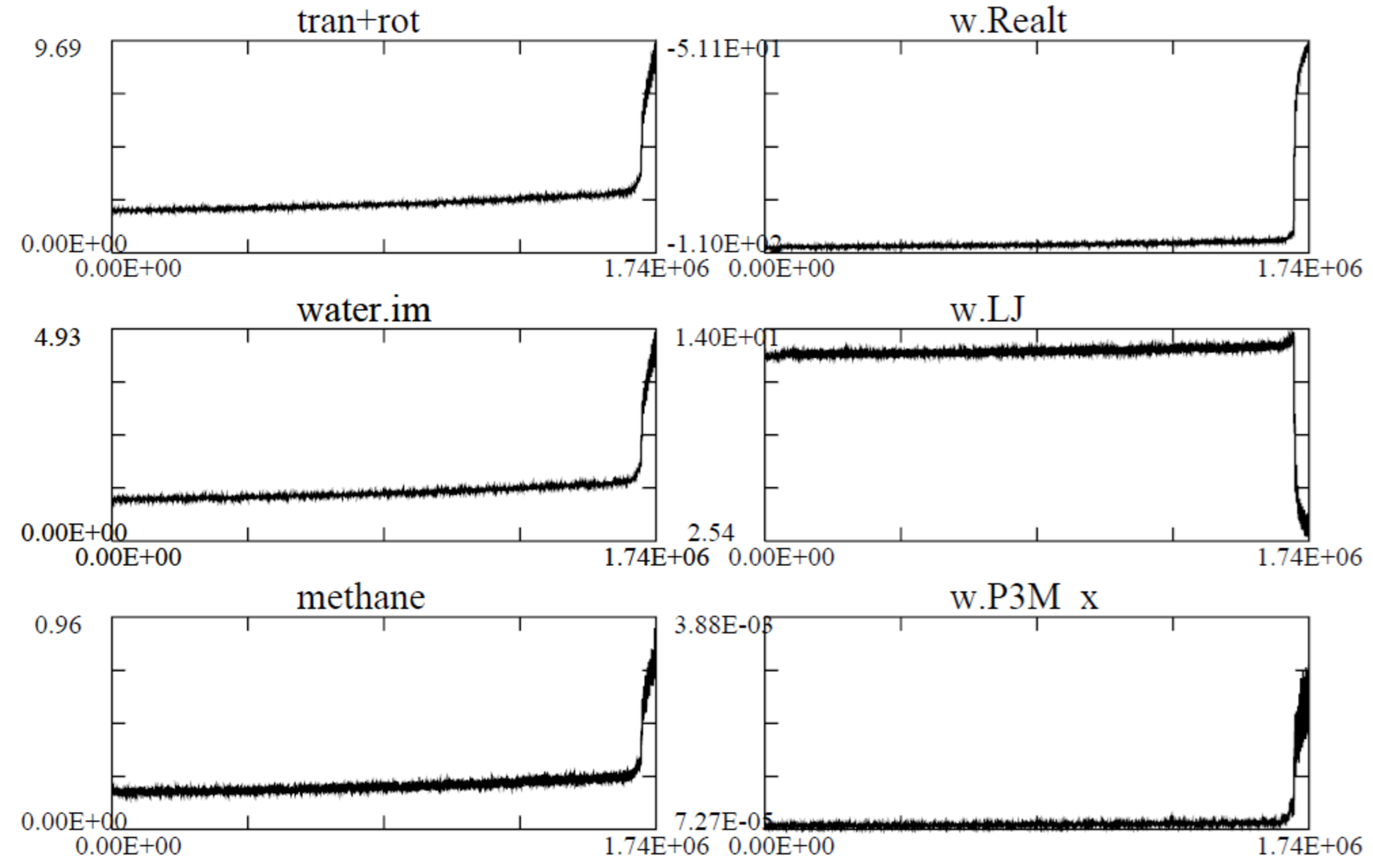} %{mhr023_105000.eps}

\caption{
Microwaves are applied to provisional methane hydrate with density of 0.912 g/cm$^3$, the initial temperature $ T= 273$ K and the external electric field $E=0.30 $ V/\AA \ 
(all the time of (0, $1.74 \times 10^{6} \tau$) including a dry run of $5 \times 10^{4} \tau$ without microwaves is shown in the figure). 
Left: The kinetic energy of translation and rotation, that of rotation, that of CH$_4$ (top to bottom, respectively). 
Right: The short-range Coulomb interaction energy, the Lennard-Jones potential energy, and P3M  energy (top to bottom, respectively).
The kinetic energy of water increases with a time rate of $\Delta W/\Delta \tau= 3.0 \times 10^{-7} W_{0}/\tau $, and the short and long-range energies eventually collapse to be liquid at the time $ t \cong 1.7 \times 10^{6} \tau$, whereas the Lennard-Jones energy decreases at the same time. 
%(The initial $E=0$ phase for $t=-5 \times 10^{4} \tau$ and 0.0 is included in Fig.\ref{tip580_Energy}).
}
\label{tip580_Energy} 
\end{figure}

\begin{figure}
 \begin{minipage}[b]{0.45\linewidth}
    \centering
    \includegraphics[keepaspectratio, scale=0.505]{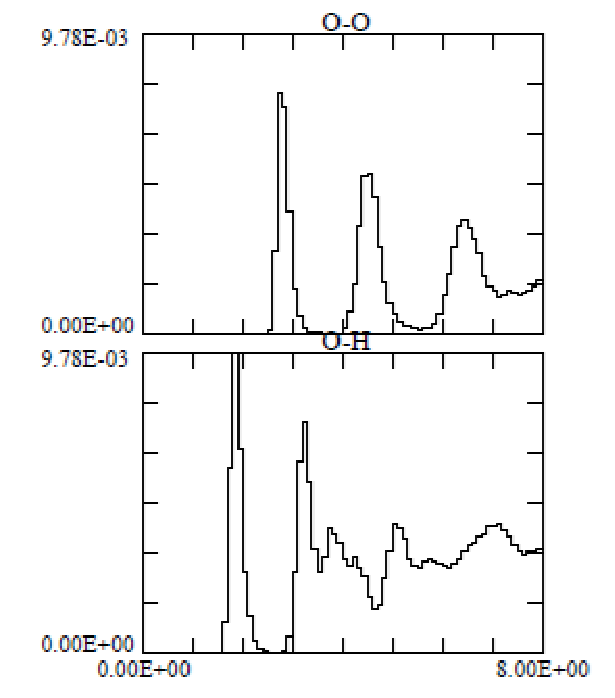}
  \end{minipage}
  \begin{minipage}[b]{0.45\linewidth}
    \centering
    \includegraphics[keepaspectratio, scale=0.505]{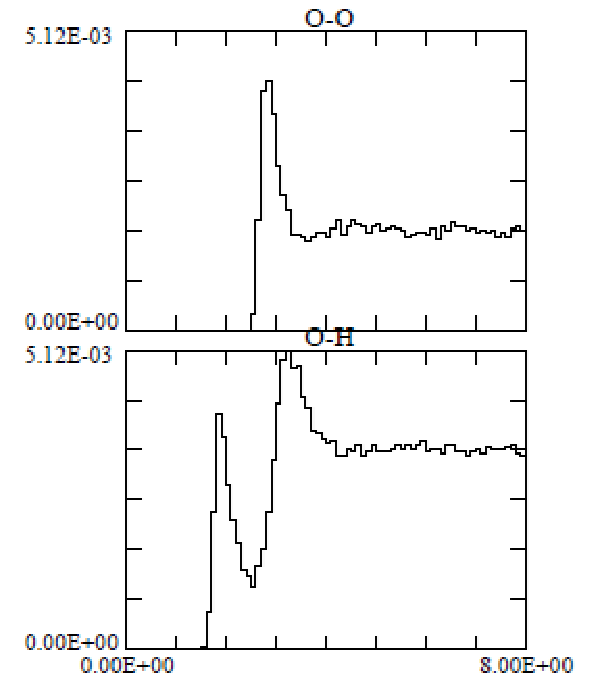}
  \end{minipage}

\caption{
The pair distribution functions between the O-O atoms (top) and O-H atoms (bottom) of the methane hydrate for the density 0.912 g/cm$^3$ and the external electric field $ E= 0.30$ V/\AA .  
The time is before the collapse at $t=1.60 \times 10^{6} \tau $ (left), % 1.65
and after the collapse at $ t =1.65 \times 10^{6} \tau $ (right). % 1.70 
%The abscissa of the distribution functions is $1.39 \times 10^{-2}$ (left), and (b) $8.75 \times 10^{-3}$ (right).
The three peaks for the O-O atoms show that the atoms have almost aggregated by the collapse in the right panel, and the two giant peaks indicate the O-H atoms.
}
\label{tip580_wat} % {Fig_FV2}
\end{figure}

\vspace{-0.3cm}
\subsection{\label{sec3.2}{Dependence of heating on the microwave fields}}

The heating of methane hydrate is done under the application of microwave fields. 
In the heating described, the volume is assumed to be constant. 
The hydrate used in this simulation is a normal pressure case (1 atm, Run A's),                  
and the density is 0.912 $\rm{g/cm}^{3}$ \cite{Sloan, density-tab}.
%                        ++++
% 0.95 $\rm{g/cm}^{3}$ for the second case \cite{matdata}, and 0.912 $\rm{g/cm}^{3}$ for the CO$_2$ hydrate in the third case. 
The run has an initial temperature of 273 K of the ice or methane hydrate. 
The translational and rotational motions are solved for the water molecules as stated in Sec.2. 
The kinetic energy of the water molecule is $W_{ekin}=(1/2)\sum_{s,j}
(M_{0}\bm{V}_{s,j}^{2} +I_{s,j} \omega_{s,j}^{2})$, where $\bm{V}_{s,j}$ and 
$\omega_{s,j}$ are the velocity and angular frequency, respectively, 
of the $s$ directions ($s=x,y,z$) and the $j$-th particle.    
$M_{0}$ is the mass of water, and $I_{s,j}$ is the inertia moment.
%A dry run that does not include any microwaves is first executed. It shows a very small and non-increasing drift, as was described at the end of Section 2. This basis is then assumed for all other runs. 

In the temperature of $T=193$ K and the pressure 1 atm, the methane hydrate undergoes a transition where the CH$_{4}$ molecules are released and vaporized as gas from the remnant cold ice.
%%The ice at the temperature 273 K has about three times less the kinetic energy compared to the methane hydrate of the same temperature and the high pressures.

\begin{figure} %[h]
\centering
  \includegraphics[width=11.0cm,clip]{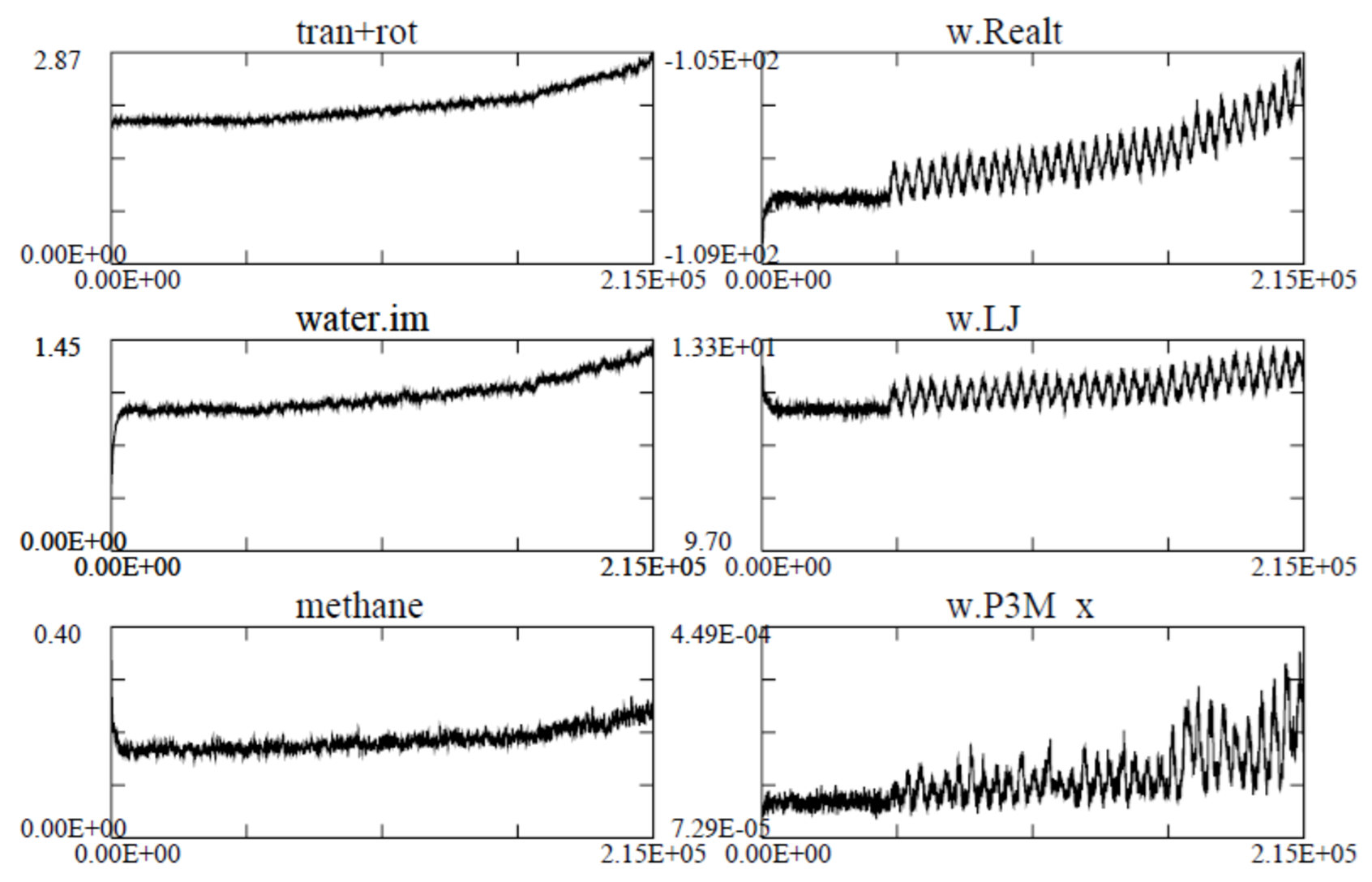} % tip600_Energy.eps
\caption{
Microwaves are applied to the ice and methane with density of 0.912 g/cm$^3$, the initial temperature $T= 273$ K and the external electric field $E=0.35 $ V/\AA \
(the time of ($0$, $2.15 \times 10^{5} \tau$) is shown for a magnificient purpose where the initial time for $5 \times 10^{4} \tau$ is a dry run without microwaves).
%% 2.15 10^5.                                                                     
Left: The kinetic energy of translation+rotation, that of rotation, that of CH$_4$ (top to bottom, respectively). 
Right: The short-range Coulomb interaction energy, the Lennard-Jones potential energy, and P3M  energy (top to bottom, respectively).
The kinetic energy of water increases with a time rate of $\Delta W/\Delta \tau= 1.6 \times 10^{-5} W_{0}/\tau$ of the linear phase $1.3 \times 10^{5} \tau$, and the short and long-range energies eventually collapse at the time $t \cong 1.8 \times 10^{5} \tau$.
%
%(The time axis starts from null in Fig.\ref{tip600_Energy}, while the $t>0$ part is executed with the microwave irradiation and the $t<0$ part is for a dry run without microwaves in the simulation.)
}
\label{tip600_Energy} 
\end{figure}

\vspace{-0.3cm}
\subsubsection{\label{sec3.2.1}{Dependence of provisional heating at the temperature 273 K}}

The methane hydrates in this subsection are simulated where they are instantaneously placed at this temperature. 
This leads to the ice and methane molecules at $T=$ 193 K or the normal and high pressure cases at $T=$ 273 K in the next subsections.

After an initial quiet phase of $t= -5 \times 10^{4} \tau$ to $t=0$ is passed, the external electric field of microwaves is applied in the $x-$direction.  
The translational and rotatinal energies and those of Coulombic and Lennard-Jones energies of water are increasing as the temperature $\Delta T= T -T_{0} > 0$ with $T_{0}=$ 273 K.
The kinetic energy of the CH$_4$ molecules increases over time, but the water and CH$_{4}$ (unified atom) molecules behave similarly in the linear phase on the left panels of Fig.\ref{tip580_scat}.
As the diagnosis of time data, it is noted that a bin size of one cycle, which has the length of $1.0 \times 10^{4} \tau$, is always chosen to minimize the periodic deviations of the heating rate in Tables \ref{Table-1} and \ref{Table-2}. 
Otherwise, it uses temporal data of that time.

The nonlinear growth of methane hydrate occurs for the time $t > 0$ of Run A1.
The increase in the energy of the water is $\Delta W/\Delta \tau \cong 3.0 \times 10^{-7} W_{0}/\tau$. 
The real dynamical system corresponds to the temperature, and the density-temperature diagram will change only for $t \gg 0$. 
The CH$_{4}$ molecules without any charges are inert to microwaves, but they interact with water molecules. The energy increases in the Lennard-Jones potential, which was close to the water case.
The temperature increase before the microwave collapse for Run A1 is 
$\Delta T \cong 61$ deg.
The total energy of the kinetic energies of water and CH$_{4}$ become very large at 
$t>1.7 \times 10^{6} \tau$, as depicted in Fig.\ref{tip580_Energy}. 
The explosive growth suddenly collapses, and the Lennard-Jones energy decreases 
chaotically.
Methane hydrate turns to be the liquid phase, as the right panel of Fig.\ref{tip580_scat}.

% T/273 = 1 particle 8.27 10^-4 /7.33 10^-4 = 1 particle W(fin)/W(start)
%$k$ is the Boltzmann constant of $1.38 \times 10^{-16} \rm{erg/K}$.   
% k= 1.38 10^{-16}, DT= W_{s}/(3726*k)= 32 deg    

The pair distribution functions of the O-O and O-H atoms are shown in 
Fig.\ref{tip580_wat} for the methane hydrate.
Their times are corresponding to before and after the collapse around $\tau \cong 1.7 \times 10^{6} \tau_{0}$.
Well separated peaks in the 8 \AA \ regions can be identified as a crystal before the collapse on the top and bottom panels of the left side, respectively. 
However, one has only one peak as liquid after the collapse, which is 
entirely buried in the $r >$ 3 \AA \ region of the O-O distribution function. Two peaks are seen with curtains in the O-H functions of the right panel. 

The heating rate of Run A2 at $E=0.35$ V/cm has an order of magnitute large in the linear phase and the run collapses earlier in $2.3 \times 10^{5} \tau$ of Table \ref{Table-1}. 
%%Run A3 collapses in short cycles of this table.

\begin{table} %[h]
\caption{The run series, fixed density ($\rm{g/cm}^3$), electric field (V/\AA), microwave heating rate$^{*}$ in averaging linear and nonlinear stages ($W_{0}/\tau$), the observed temperature $T$ and time, and guest molecules. The volume of simulations are different from the high density in Run B1 to the medium density Run B3, and the normal density  as the reference Run A2R.
The linear stage of the heating rate is shown in these runs. 
(The time axis in the abscissa starts from null, while in the simulation the $t>0$ part is executed with the microwave irradiation for Table \ref{Table-2}$^{*}$.)
}
%%
%% a time rate $\Delta W_{dr}/\Delta \tau= 1.6 \times 10^{-7} W_{0}/\tau$ 
%% this is a baseline
\label{Table-2}
  \centering 
  \vspace{-0.2cm}
  \begin{tabular}{clllllll} 
\hspace{0.3cm} \\ \hline
 \hspace{0.2cm} series & density & $E$ & heating rate$^{*}$ & observed $T$ and time & status & guest \hspace{0.1cm} \\ \hline \hline
 B1$^{*}$ & $0.963 \rm{g/cm}^3$ & 0.35 V/\AA & $1.6 \times 10^{-7} W_{0}/\tau$ & 280. \ $3.4 \times 10^{5}\tau$ & running & $\rm{CH}_4$ 
%  E=0 average, then start
 \\
% 1.6 \times 10^{-7} as the zero beseline
 B2$^{*}$  & $0.941 \rm{g/cm}^3$ & 0.35 V/\AA & $ 2.0 \times 10^{-7} W_{0}/\tau$ & 282. \ $3.4 \times 10^{5}\tau$ & running & $\rm{CH}_4$ 
% same above 
 \\
 B3$^{*}$ & $0.930 \rm{g/cm}^3$ & 0.35 V/\AA & $4.4 \times 10^{-7} W_{0}/\tau$ & 300. \ $4.5 \times 10^{5}\tau$ & running & $\rm{CH}_4$ 
% negative -> positive, E=0 -> ave=0.0, 
 \\ 
%         ++++
 A2R$^{*}$ & $0.912 \rm{g/cm}^3$ & 0.35 V/\AA & $2.8 \times 10^{-6} W_{0}/\tau$ & 318. \ $1.2 \times 10^{5}\tau$ & collapsed & $\rm{CH}_4$
% \\
% linear: 17.0 318.
% B4 & $0.866 \rm{g/cm}^3$ & 0.35 V/\AA & $ 6.5 \times 10^{-6} W_{0}/\tau$ & $ 0.8 \times 10^{4}\tau$ \ collapsed & $\rm{CH}_4$ % 5.5
%
\hspace{0.3cm} \\ \hline
  \end{tabular}
\end{table}

\begin{table} %[h]  %            ++++++
\caption{The run series of Run C1 in the initial temperature 193 K, fixed density ($\rm{g/cm}^3$), electric field (V/\AA), microwave heating rate$^{*}$ in averaging linear and nonlinear stages ($W_{0}/\tau$) with $W_{0}$ the initial kinetic energy of methane hydrate, the observed temperature $T$ and time for the run, and guest molecules.
(The time axis in the abscissa starts from null, while the $t>0$ part is executed with the microwave irradiation in the simulation for Table \ref{Table-3}$^{*}$.)
}
\label{Table-3}
  \centering 
  \vspace{-0.2cm}
\begin{tabular}{cllllll} 
\hspace{0.3cm} \\ \hline
\hspace{0.2cm} series & density & $E$ & heating rate$^{*}$ & observed $T$ and time & status & guest \hspace{0.1cm} \\ \hline \hline
%++              ++++
 C1$^{*}$ & $0.912 \rm{g/cm}^3$ & 0.35 V/\AA & $5.9 \times 10^{-7} W_{0}/\tau$ & 212. \ $ 1.6 \times 10^{5} \tau$ & running &  $\rm{CH}_4$
 \hspace{0.3cm} \\ \hline  
  \end{tabular}
\end{table}

\vspace{-0.3cm}
\subsubsection{\label{sec3.2.2}{Dependence of heating at the temperature 273 K}}

The ice but without $\rm{CH}_{4}$ are simulated to look at the subtle molecular dynamics simulation at the temperature 273 K in Run A0R and Run A1R. %in this subsection.
%         without methane
In the initial time $t= -5 \times 10^{4}\tau$ to $t= -5 \times 10^{4}\tau +1 \times 10^{3} \tau$, the ice molecules make dry runs. %methane molecules are free from the cages of methane hydrate sites. Then, the ice cages are closed after that time. 
The heating rate in the linear phase is $4.5 \times 10^{-8} W_{0}/\tau$ for $E=$ 0.10 V/\AA \ of Run A0R.
The heating rate at for $E=$ 0.30 V/\AA \ of Run A1R is $1.8 \times 10^{-7} W_{0}/\tau$ at $ t \cong 3.7 \times 10^{5} \tau$ in the linear phase. 
It goes into the nonlinear phase around  $ t > 6.5 \times 10^{5} \tau$, and approches the collapse phase aronnd $t \cong 9.6 \times 10^{5} \tau$ and $dT \cong 60$ deg.
But, the water molecules are not disturbed until $dT \cong 120$ deg because $\rm{CH}_4$ is purposely not included to see the subtle molecular dynamics for Run A0R and Run A1R, as the lower part of Table \ref{Table-1}.

One notices that after $t=0$, the kinetic energy with $E= 3.5 \times 10^{7} \rm{V/cm}$ in Run A2R of Table \ref{Table-1} drastically increases, and the heating rate at the linear stage is $\Delta W/\Delta \tau= 2.8 \times 10^{-6} W_{0}/\tau$ at the time $ t \cong 1.2 \times 10^{5} \tau$ in Fig.\ref{tip600_Energy}.
The ice continues to the nonlinear stage with $\Delta T \cong 73$ deg at $1.6 \times 10^{5} \tau$ and the collapse occurs at $t \cong 2.2 \times 10^{5} \tau$.
%Also, the simulation of the density of 0.7950 g/cm$^{3}$ without methane molecules is almost the same result with that of the free methane molecules.

%%Another simulation with the electric field $E= 4.0 \times 10^{7} \rm{V/cm}$ is executed in Run A3R. % new A3R
%%The heating rate becomes very large which is $\Delta W/\Delta \tau \cong 1.7 \times 10^{-5} W_{0}/\tau$ and $T \cong 333$ deg in the linear phase $ t \cong 2.5 \times 10^{4} \tau$.
%%Then, it goes to the explosive phase and collapses at $ t \cong 2.9 \times 10^{4} \tau$.

\vspace{-0.1cm}
\subsection{\label{sec3.3}Dependence of heating on high and medium densities}

Methane hydrate of the high density of $0.963 \rm{g/cm}^{3}$ with increased pressure is started from a new initial condition as Run B1 of Table \ref{Table-2}.   
The tilted baseline in $E=0$ has been subtracted from the Run B1 value.
%The E=0 bin is averaged as the baseline                         ++++++
%
This high density has been continued where the kinetic energy by microwave heating increases for the small amount by $1.6 \times 10^{-7} W_{0}/\tau$ at the time of $3.1 \times 10^{5} \tau$.
The mediumly high density of $0.941 \rm{g/cm}^{3}$ as Run B2 is also stable whose heating rate is $2.0 \times 10^{-7} W_{0}/\tau$ in the TIP5P-Ewald model simulation.
The temperature increase is not small by the microwave irradiations, but is finite for about 10 deg in $3 \times 10^{5} \tau$ for Run B1 and Run B2. 
%We may then put methane hydrate to the nonlinear and collapse phase in ten times more simulations. 

%         Run A2R
For the medium density of $0.930 \rm{g/cm}^{3}$ in Run B3, i.e. the mid-point between Run B2 and Run A2R as a reference, methane hydrate seems to be stable without the microwave irradiation $E=0$ for the time $t < 0$.  
The heating rate with the microwave at $E=0.35$ V/\AA \ becomes $3.5 \times 10^{-7} \rm{W}_{0}/\tau$ and $\Delta T \cong$ 10 deg at $t = 2 \times 10^{5} \tau$.  
Then after this time, it goes in the nonlinear stage which increases the heating rate as listed in Table \ref{Table-2}. 
%                                               ice
However, the normal density of the ice $0.912 \rm{g/cm}^{3}$ in Run A2R was going unstable and collapsed after the microwave heating as shown in Tables \ref{Table-1} and \ref{Table-2}.
%
%For the low density of $0.87 \rm{g/cm}^{3}$ with low pressure case in Run B4, the short heating is observed which is followed by the collapse of methane hydrate.
% in Fig.\ref{tip575_Energy}.  

About the five-body water model in Run A1, it is shown that there is a factor of two the difference of the collapse time of the TIP5P-Ewald model from the short collapse time of the SPC/E model in the same parameters \cite{TSN}.

\begin{figure}
\centering
  \includegraphics[width=11.1cm,clip]{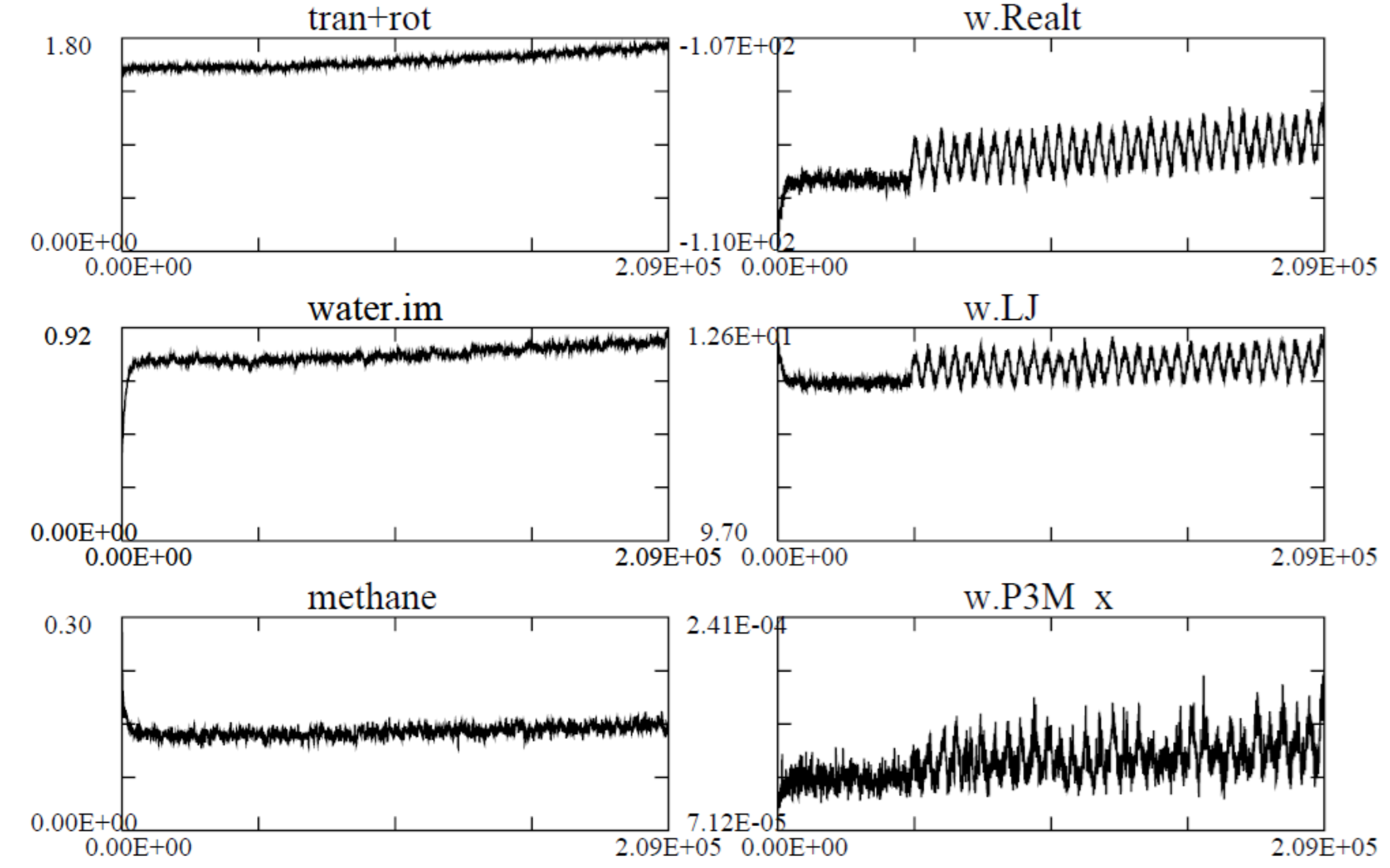}  %[width=7.1cm,clip]
%                                                                                            +++++
\caption{Microwaves are applied to methane hydrate with density of 0.912 g/cm$^3$, the initial temperature $T= 193$ K and the external electric field $E=0.35 $ V/\AA \ as Run C1
(the time from $t=0$ to $5 \times 10^{4} \tau$ for a dry run without microwaves is plotted in the figure).
Left: The kinetic energy of translation+rotation, that of rotation, that of CH$_4$ (top to bottom, respectively). 
Right: The short-range Coulomb interaction energy, the Lennard-Jones potential energy, and P3M  energy (top to bottom, respectively).
(The time axis in the abscissa starts from null in Fig.\ref{tip530_Energy}, while the $t>0$ part is executed with the microwave irradiation in the simulation.) 
}
\label{tip530_Energy}  % tip575_Energy
\end{figure}

\vspace{-0.3cm}
\begin{figure}
\centering
  \includegraphics[width=8.1cm,clip]{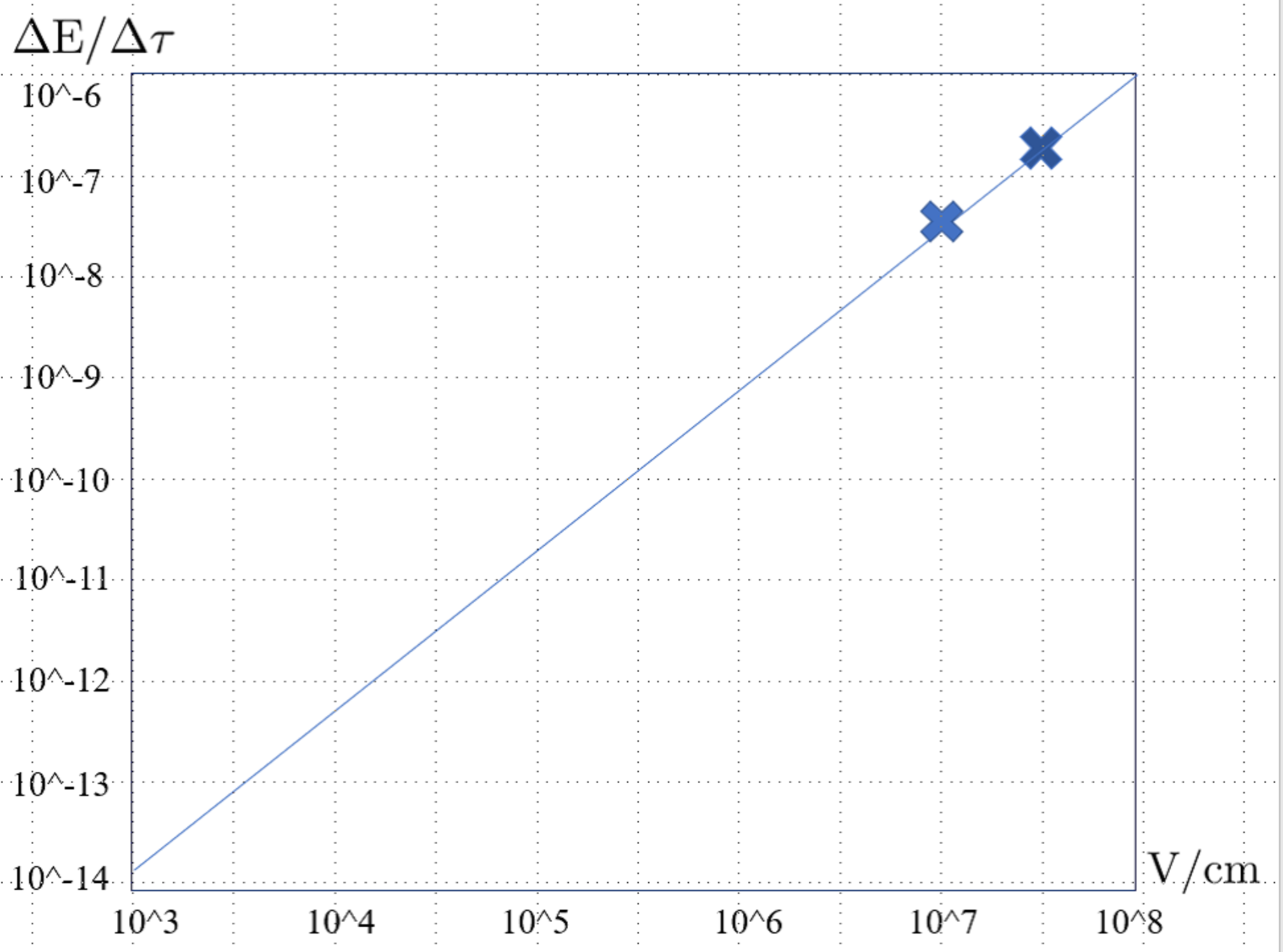} %[width=7.1cm,clip]
\caption{The time derivative of the kinetic energy for the external electric field from $E=$ 0.10 V/\AA\ (Run A0R) to $E=$ 0.30 V/\AA\ (Run A1R) is extrapolated against the run at $E= 1 \times 10^{-5}$ V/\AA\ = 1,000 V/cm. One finds the most left-side point of the figure.}

\label{tip600_EkScale} %tip580_EtotScale}
\end{figure}

\subsection{\label{sec3.4}{Dependence of heating on the temperature 193 K}}

The initial temperature 193 K with the pressure 1 atm is simulated as Run C1 of Table \ref{Table-3}.
It is the methane hydrate regime at this temperature before the transition occurs, but is not known when the ice and methane molecules would overcome the former.  It is assumed that the methane hydrate is there for the current picture. 

%The temporal drift kinetic energy which is minimized at $t= 3 \times 10^{4} \tau$ in  microwave irradiation is subtracted. 
The methane hydrate is stable without microwave irradiation.
For microwaves after $t > 0$, the kinetic energy is increased and the heating rate is $\Delta W/\Delta \tau= 6.7 \times 10^{-7} W_{0}/\tau$ and the temperature rise is $\Delta T \cong 17 $ deg at $t= 1.6 \times 10^{5} \tau$.

One should assume, however, that the temperature should be constant during the transition phase.
Because the macroscale effect is not included in the molecular dynamics simulations, the methane hydrate and ice transition must be studied separately \cite{Trans}.

\subsection{\label{sec3.5}The microwave devices of the input power 1,000 V/cm}

The normal input power of the microwave devices is estimated. 
The time derivative of the kinetic energy of Run A0R - Run A1R is plotted against the microwave devices of $1 \times 10^{3}$ V/cm= $1 \times 10^{-5}$ V/\AA \ in Fig.\ref{tip600_EkScale}.
The data of $\Delta E_{kin}/\Delta \tau= 2.3 \times 10^{-7} W_{0}/\tau$ of the linear phase of Run A1R is reduced to $1.0 \times 10^{-14} W_{0}/\tau$, which is dynamically unstable to the most left point of the figure. 
The conversion from 10 GHz to 2.45 GHz together becomes the final collapse time as $(1.3 \times 10^{6} \times 10^{-14} \rm{s}) \times (2.3 \times 10^{-7}/1.1 \times 10^{-14}) \times (10/2.45) \cong$ 1.1 s when the 100\% microwave efficiency is applied.

\vspace{-0.3cm}
\section{\label{sec4}Summary}

Methane hydrate was simulated by molecular dynamics with periodic boundary conditions. The normal density of 0.91 g/cm$^3$ represented the sea level conditions, and the density of 0.96 g/cm$^{3}$ corresponded to the high-pressure condition at 273 K. 
% two digits: 0.91 g/cm^3
The microwave electric field was applied while the volume was held constant. 

The provisional simulation of the methane hydrate in the temperatures of 273 K and the pressure 1 atm was executed.
%, the methane was escaped from the CH$_{4}$-free hydrate.
For Run A1 with the external electric field $E= 0.30$V/\AA  , the heating rate was $3.0 \times 10^{-7} W_{0}/\tau$.
The unstable methane hydrate collapsed as liquid for the pressure 1 atm,  
and the temperture increase was $\Delta T \cong 61 $ deg at $ t \cong 1.7 \times 10^{6} \tau $. 
%                                                               +++++++++++++++++
%In the external electric field $E=$ 0.35 V/\AA , the methane hydrate collapsed at the time $t \cong 2.3 \times 10^{5}\tau$. %
%The methane-free ice at the temperature of 273 K was simulated in Run A2R, whose heating rate was larger and collapsed around the same periods as that of Run A2 in the lower part of Table \ref{Table-1}.
%%For Run A3 of $E= 0.40$ V/\AA , the heating rate was very large as $ 1.6 \times 10^{-5} W_{0}/\tau$, and methane hydrate collapsed shortly at $t \cong 1.3 \times 10^{4} \tau$. % 6.3

In the ice but without $\rm{CH}_{4}$ to see the molecular dynamics, the heating rate in the linear phase starting $T=$ 273 K was $4.1 \times 10^{-8} W_{0}/\tau$ for $E=$ 0.10 V/\AA \ and that for $E=$ 0.30 V/\AA \ was $2.3 \times 10^{-7} W_{0}/\tau$, both of which were in the linear phase at Run A0R and Run A1R, respectively.
%That of the collapsed simulation was $\Delta W/\Delta \tau= 2.8 \times 10^{-6} W_{0}/\tau$ at the time $ t \cong 1.2 \times 10^{5} \tau$ for Run A2R.
%%, and that was $\Delta W/\Delta \tau \cong 1.7 \times 10^{-5} W_{0}/\tau$ at $ t \cong 2.5 \times 10^{4} \tau$ for Run A3R.

On the other hand, for the high density of $0.963 \rm{g/cm}^3$ in RunB1 and also $0.941 \rm{g/cm}^{3}$ in Run B2 for the increased pressures, the kinetic energy remained stable and the small microwave heating was obseved.
The kinetic energy of the medium density $0.930 \rm{g/cm}^{3}$ in Run B3 seemed stable without microwaves, but it increased nonlinearly with the microwave irradiation.
The stable and unstable boundary at the temperature 273 K without microwave appeared to be around $0.930 \rm{g/cm}^{3}$, which was pointed to the pressure 2.3 MPa and temperature 273 K.

The microwave heating of methane hydrate on the temperature 193 K and the pressure 1 atm was argued before the transition was reached in Run C1.

The ice in the devices of the input $E=$ 1,000 V/cm and the pressure 1 atm was  collapsed about 1 s when the 100\% microwave efficiency was applied.

%\newpage
\vspace{0.7cm}
\noindent
{\Large {\bf Acknowledgments}}

\vspace{0.2cm}
The author is grateful to Dr. M. Matsumoto of Okayama University for the initial configurations of the ice and hydrate structures. He thanks Dr. Y. Zempo for collaborations on the classical molecular dynamics code and the spectrum analysis code for the maximum entropy method. 
Also, he thanks Dr. S. Nakatani and Dr. M. Sato for useful discussions of methane hydrate experiments.  
The computation is performed by the NEC SX-Aurora TSUBASA and the Intel LX Server of National Institute of Fusion Science  in Toki City, Japan.

%\newpage

    \renewcommand{\theequation}{A.\arabic{equation}}
    \setcounter{equation}{0}

\vspace{0.7cm}
\noindent
{\Large \label{append-A}
{\bf Appendix: Long-range Coulombic interactions}
}

\vspace{0.2cm}
The long-range Coulombic interactions $\bm{F}_{LR}(\bm{r})$ are defined in the Fourier space 
using $dn(n_{\gamma})$, $G(n_x,n_y,n_z)$, $\bm{K}(n_x,n_y,n_z)$ and $\Delta(n_x,n_y,n_z)$ functions
in Eq.(\ref{Longrg1}), Eq.(\ref{Longrg}), which are written in the periodic boundary conditions,  
\begin{equation}
   G(n_x,n_y,n_z) = (2 M_x M_y M_z/L^{2}) \times
   \Bigl[ dn(n_x)K_x +dn(n_y)K_y +dn(n_z)K_z \Bigr] 
  /(\Lambda \Delta^{2}), 
\end{equation}
\begin{eqnarray}
  && \hspace*{-0.3cm}
   \bm{K}(n_x,n_y,n_z) = \sum_{n_{1},n_{2},n_{3}} 
   (n_1,n_2,n_3) \Bigl\{ \exp \Bigl (-\Bigl( \pi/(\alpha L)  \Bigr)^{2} \Bigr) /\Lambda 
   \Bigr\} \times \nonumber \\
   && \hspace*{1.1cm} 
   \Bigl( sinc \Bigl( \frac{n_x+M_x n_1}{M_x} \Bigr) \Bigr)^{2P} 
   \Bigl( sinc \Bigl( \frac{n_y+M_y n_2}{M_y} \Bigr) \Bigr)^{2P} 
   \Bigl( sinc \Bigl( \frac{n_z+M_z n_3}{M_z} \Bigr) \Bigr)^{2P}   
\end{eqnarray}
\begin{eqnarray}
   && \hspace*{-1.3cm}
   \Delta(n_x,n_y,n_z) = \sum_{n_{1},n_{2},n_{3}} 
   \Bigl( sinc \Bigl( \frac{n_x +M_x n_1}{M_x} \Bigr) \Bigr)^{2P} \times
   \nonumber \\
   && \hspace*{3.1cm}
   \Bigl( sinc \Bigl( \frac{n_y +M_y n_2}{M_y} \Bigr) \Bigr)^{2P} 
   \Bigl( sinc \Bigl( \frac{n_z +M_z n_3}{M_z} \Bigr) \Bigr)^{2P}, 
\end{eqnarray}
\begin{eqnarray}
  && \Lambda= dn(n_{x})^2 +dn(n_{y})^2 +dn(n_{z})^2.  
\label{Longrg2}
\end{eqnarray}

The first Brillouin zone should take the summation of $-1 \leq n_{1} \leq 1$ (the degree is $P=3$), and M$_x$ is the mesh in the $x-$direction; the same procedures should be taken in the other directions due to the tetragonal crystal symmetry.
The sinc function is used to account for the long slopes of the $\bm{K}(n_x,n_y,n_z)$ and $\Delta(n_x,n_y,n_z)$ functions.
The index ranges for the $G(n_x,n_y,n_z)$ function are $0 \le n_x \le M_x/2$, $0 \le n_y \le M_y-1$ and $0 \le n_z \le M_z-1$, where M$_x$, M$_y$, and M$_z$ are the number of points in the $x, y$, and $z$ directions, respectively.

%\newpage
%\vspace{0.5cm}

\end{document}